\documentclass[Times,4pt,aps,pra,twocolumn,showpacs,amsmath,amssymb,floatfix,footinbib,superscriptaddress,times]{revtex4-1}
\usepackage{amsmath,natbib,graphicx,subfigure,amssymb,graphics,amsmath,mathrsfs,CJK,color}
\usepackage{multirow,fancyhdr,color,bm,tabularx,psfrag,dcolumn,datetime}
\usepackage[dvipdfm,colorlinks=true,linkcolor=blue,urlcolor=blue,citecolor=blue]{hyperref}
\usepackage[ansinew]{inputenc}
\usepackage[usenames,dvipsnames]{pstricks}
\usepackage[pagewise,mathlines]{lineno}
\usepackage{tikz,xcolor,hyperref}
\definecolor{lime}{HTML}{A6CE39}
\DeclareRobustCommand{\orcidicon}{%
    \begin{tikzpicture}
    \draw[lime, fill=lime] (0,0)
    circle [radius=0.16]
    node[white] {{\fontfamily{qag}\selectfont \tiny ID}};\draw[white, fill=white] (-0.0625,0.095)
    circle [radius=0.007];
    \end{tikzpicture}
    \hspace{-2mm}}
\foreach \x in {A, ..., Z}
{\expandafter\xdef\csname orcid\x\endcsname{\noexpand\href{https://orcid.org/\csname orcidauthor\x\endcsname}{\noexpand\orcidicon}}}

\begin{document}
\title{Quantum dynamics evolution predicted by the long short-term memory network in the photosystem II reaction center}

\author{Zi-Ran Zhao}
\affiliation{Faculty of science, Kunming University of Science and Technology, Kunming, 650093, PR China}
\affiliation{Center for Quantum Materials and Computational Condensed Matter Physics, Faculty of Science, Kunming University of Science and Technology, Kunming, 650500, PR China}
\author{Shun-Cai Zhao\orcidA{}}%
\email[Corresponding author: ]{zhaosc@kmust.edu.cn.}
\affiliation{Faculty of science, Kunming University of Science and Technology, Kunming, 650093, PR China}
\affiliation{Center for Quantum Materials and Computational Condensed Matter Physics, Faculty of Science, Kunming University of Science and Technology, Kunming, 650500, PR China}
\author{Yi-Meng Huang}
\affiliation{Faculty of science, Kunming University of Science and Technology, Kunming, 650093, PR China}
\affiliation{Center for Quantum Materials and Computational Condensed Matter Physics, Faculty of Science, Kunming University of Science and Technology, Kunming, 650500, PR China}
\date{\currenttime,~\today}

\begin{abstract}
Predicting future physical behavior from limited theoretical simulation data is an emerging research paradigm driven by the integration of artificial intelligence and quantum physics. In this work, charge transport (CT) behavior was predicted over extended time scales using a deep learning model-the long short-term memory (LSTM) network with an error-threshold training method-in the photosystem II reaction center (PSII-RC). Theoretical simulation data within 8 fs were used to train the modified LSTM network, yielding distinct predictions with differences on the order of $10^{-4}$  over prolonged periods compared to the training set collection time. The results highlight the potential of LSTM to uncover the underlying physics governing CT beyond conventional quantum physical methods. These findings warrant further investigation to fully explore the scope and efficacy of LSTM in advancing our understanding of photosynthesis at the molecular scale.(\href{https://pan.baidu.com/s/1nbQsKLFoplun5AaTJOON1A?pwd=zzrz}{Original codes in Supplement information})
\end{abstract}

\maketitle
\tableofcontents

\section{Introduction}

Accurately modeling quantum open systems remains fundamentally challenging due to complex interactions between a target system and its high-dimensional, strongly coupled environment~\cite{PhysRevLett.103.210401,Vacchini2011}. While the system-plus-bath framework prioritizes the reduced dynamics of the target system~\cite{shibata1977reduced,rivas2012open}, existing theoretical approaches struggle to balance computational efficiency with predictive accuracy across diverse parameter regimes.

Perturbative methods (e.g., Redfield theory~\cite{purkayastha2021beyond,stace2021redfield,trushechkin2021redfield}) fail under strong system-bath coupling. Numerically exact approaches-hierarchical equations of motion (HEOM)~\cite{tanimura2020efficient,wiedmann2020modular,tang2021hierarchical,erpenbeck2021hierarchical}, tensor-network decompositions~\cite{chin2020tensor}, and multi- configurational techniques\cite{tang2020hybrid,wang2020multi,farag2021multi}-resolve this limitation but incur prohibitive computational costs or convergence issues~\cite{schuch2019convergence,chen2021overcoming,schroder2022multi}. Machine learning (ML) offers alternative solutions~\cite{schmidt2019machine,elton2019deep,dunjko2018machine,ntampaka2019machine}, yet standard architectures lack temporal resolution for quantum dissipative dynamics~\cite{palmieri2020quantum}, and their accuracy critically depends on training data quality~\cite{rodrigues2020artificial,carleo2017solving}.

We propose a long short-term memory neural network (LSTM-NN) framework~\cite{zhu2022long} optimized for charge-transfer (CT) dynamics in photosystem II reaction centers (PSII-RC). By designing task-specific gate architectures (forget/input/output gates) and layer configurations, our model achieves high-precision long-term predictions using minimal training data from quantum master equations-surpassing conventional ML implementations~\cite{taieb2018forecasting,lai2018modeling}. Unlike feed-forward networks, LSTM's recurrent connections inherently capture long-range temporal correlations~\cite{garciaperez2021machine,potocnik2022forecasting}, addressing gradient vanishing/exploding issues~\cite{lim2021review} while reducing data dependency.

The structure of this paper is as follows: Section II describes the theoretical framework and LSTM-NN architecture optimization. Section III presents the results and discusses the model's performance benchmarking against exact solutions. Finally, Section IV summarizes our findings and outlines future directions.

\section{Dataset collection from Photosystem II reaction center (PSII-RC)}
\subsection{Theoretical model}
\begin{figure}
\centering
\subfigure[]{\includegraphics[width=0.40\columnwidth]{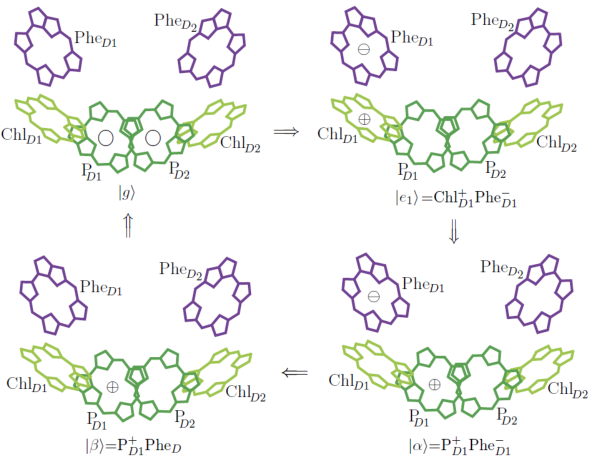}}
\hspace{0.20in}
\subfigure[]{\includegraphics[width=0.45\columnwidth]{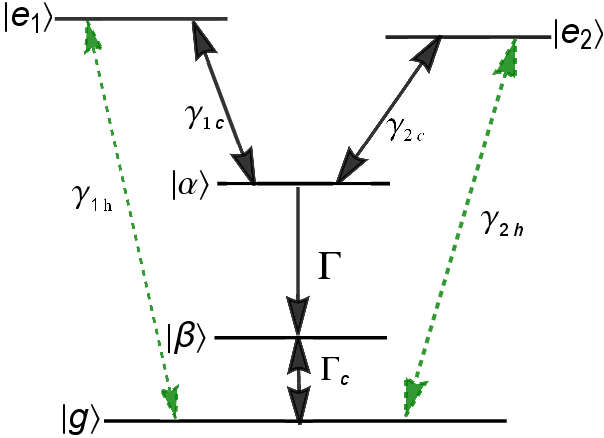}}
\caption{(a) Four typical sates in the charge-transfer process\cite{author2021charge}. On the channel of $|g\rangle$$\rightarrow$$|e_{1}\rangle$$\rightarrow$$|\alpha\rangle$$\rightarrow$$|\beta\rangle$$\rightarrow$$|g\rangle$) of the six core-pigments in the PSII RC. \(|g\rangle\), six pigments in neutral ground state. \(|e_{1}\rangle\)=Chl\(^{+} _{D1}\)Phe\(_{D1}^{-}\) after Chl\(_{D1}\) primary charge-transfer state when the electron donor rapidly loses an electron to the nearby electron acceptor molecule Phe\(_{D1}\). \(|\alpha\rangle\)=P\(_{D1}^{+}\)Phe\(_{D1}^{-}\) charge-transfer state, after the positive and negative charges are spatially separated. \(|\beta\rangle\)=P\( _{D1}^{+}\)Phe\( _{D1}\), positively charged state, after an electron has been released from the system to perform work. (b) Corresponding five-level diagram of the photosynthetic QHE model with two photon-absorbing channels $|g\rangle$ to $|e_{1}\rangle$ and $|e_{2}\rangle$. $\gamma_{1h}$ and $\gamma_{2h}$ are the transition rates corresponding, $|\alpha\rangle$ is the charge separation state characterized by the transition rate $\gamma_{1h}(\gamma_{2h})$ from state $|e_{1}\rangle(|e_{2}\rangle)$ to state $|\alpha\rangle$, and the relaxation rate $\Gamma$ to state $|\beta\rangle$, $\Gamma_{C}$ is the relaxation rate to the ground state $|g\rangle$ of the system.}\label{Fig.1}
\end{figure}

The PSII-RC in purple bacteria and oxygen-evolving organisms contains six pigments aligned in two branches\cite{M2017Effects}, consisting of four chlorophylls (special pair P\(_{D1}\) and P\(_{D2}\), accessory Chl\(_{D1}\) and Chl\(_{D2}\)) and two pheophytins (Phe\(_{D1}\) and Phe\(_{D2}\)) \cite{2015Multiple,author2021charge}. As shown in Fig.\ref{Fig.1} (a), \(|g\rangle\) represents the neutral ground state of all six pigments, \(|e_{1}\rangle\)=Chl\(^{+} _{D1}\)Phe\(_{D1}^{-}\)  denotes the primary charge-transfer state, $|\alpha\rangle$ =P\(_{D1}^{+}\)Phe\(_{D1}^{-}\) describes the charge-separated state, and \(|\beta\rangle\) = P\( _{D1}^{+}\)Phe\( _{D1}\) represents the positively charged state after the electron release.

The PSII-RC system is modeled as a five-level quantum heat engine (QHE) with photon-absorbing transitions from $|g\rangle$ to $|e_{1}\rangle$ and $|e_{2}\rangle$ at rates $\gamma_{1h}$ and $\gamma_{2h}$, respectively, and charge-separation processes $|e_{1}\rangle$$\leftrightarrow$$|\alpha\rangle$$\leftrightarrow$$|\beta\rangle$, $|e_{2}\rangle$$\leftrightarrow$$|\alpha\rangle$$\leftrightarrow$$|\beta\rangle$ occurring at rates  $\gamma_{1c}$ and $\gamma_{2c}$ (Fig.\ref{Fig.1} (b)). The total Hamiltonian is:

\begin{align}
\hat{H}=\hat{H}_{S}+\hat{H}_{B}+\hat{H}_{SB}, \label{eq1}
\end{align}

\noindent where the system Hamiltonian $\hat{H}_{S}$  is given by:
\begin{eqnarray}
& \hat{H}_{S}=&E_{0}|g\rangle\langle g|+E_{e_{1}}|e_{1}\rangle\langle e_{1}|+E_{e_{2}}|e_{2}\rangle\langle e_{2}|\nonumber\\
&&+E_{\alpha}|\alpha\rangle\langle \alpha|+E_{\beta}|\beta\rangle\langle \beta|,  \label{eq2}
\end{eqnarray}

\noindent Among Eq.(\ref{eq1}), $\hat{H}_{B}$ representing the ambient thermal environment, is modeled as a harmonic oscillator bath:
\begin{align}
\hat{H}_{B}=\sum_{k}\hbar\omega_{k}\hat{a}^{\dag}_{k}\hat{a}_{k}, \label{eq3}
\end{align}

\noindent where $\hat{a}^{\dag}_{k}(\hat{a}_{k})$ are the creation (annihilation) operator of the \emph{k}-th harmonic oscillator mode with its frequency \(\omega_{k}\). In Eq.(\ref{eq1}),
the interaction Hamiltonian $\hat{H}_{SB}$, under the rotating-wave approximation, is expressed as:
\begin{align}
&\hat{H}_{SB}=\hat{V}_{h} + \hat{V}_{c}\nonumber, \\
&\hat{V}_{h}=\sum_{i=1,2}\sum_{k}\hbar(\varepsilon_{ik}\hat{\sigma}_{gi}\otimes\hat a_{k}^{\dag}+\varepsilon_{ik}^{\ast}\hat{\sigma}_{gi}^{\dag}\otimes\hat a_{k})\nonumber, \\
&\hat{V}_{c}=\sum_{j=1,2}\sum_{k}\hbar(\varepsilon_{jk}\hat{\sigma}_{\alpha j}\otimes\hat b_{k}^{\dag}+\varepsilon_{jk}^{\ast}\hat{\sigma}_{\alpha j}^{\dag}\otimes\hat b_{k}).  \label{eq4}\\\nonumber
\end{align}

\noindent where $\varepsilon_{ik}(\varepsilon_{jk})$ is the corresponding coupling-strength between the i(j)th\(_{(i,j=1,2)}\) charge transfer process and the kth mode of ambient environment reservoir, and the transit operators are defined as $\hat{\sigma}_{gi}=|g\rangle\langle e_{i}|$, $\hat{\sigma}_{\alpha j}$=$|\alpha\rangle\langle e_{j}|_{(i,j=1,2)}$ with $\hat{a}^{\dag}_{k}$, $\hat{b}^{\dag}_{k} (\hat{a}_{k}$, $\hat{b}_{k})$ being the creation(annihilation) operators of the \emph{k}-th reservoir mode, respectively.

Under the Born-Markov and Weisskopf-Wigner approximations in the Schr$\ddot{o}$dinger picture, the PSII RC can be described by the master equations as follows,

\begin{equation}
\frac{d\hat{\rho}}{dt}=-i[\hat{H}_{S},\hat{\rho}]+\sum_{i,j=1,2}\mathscr{L}_{ijh}\hat{\rho}+\mathscr{L}_{ijc}\hat{\rho}+\mathscr{L}_{\Gamma_{C}}\hat{\rho}+\mathscr{L}_{\Gamma}\hat{\rho}, \label{eq5}
\end{equation}

\noindent In Eq.(\ref{eq5}), the first term on the right side describes the coherent evolution of the PSII RC, and the other four Lindblad-type superoperator terms are deduced with the following expressions,

\begin{widetext}
\begin{eqnarray}
&\mathscr{L}_{ijh}\hat{\rho}=&\sum_{i,j=1,2}\frac{\gamma_{ijh}}{2}[(n_{ih}+1)(\hat{\sigma}_{gi}\hat{\rho}\hat{\sigma}_{gj}^{\dag}+\hat{\sigma}_{gj}\hat{\rho}\hat{\sigma}_{gi}^{\dag}
-\hat{\sigma}_{gj}^{\dag}\hat{\sigma}_{gi}\hat{\rho}-\hat{\rho}\hat{\sigma}_{gj}^{\dag}\hat{\sigma}_{gi})\nonumber\\
&&+n_{ih}(\hat{\sigma}_{gi}^{\dag}\hat{\rho}\hat{\sigma}_{gj}+\hat{\sigma}_{gj}^{\dag}\hat{\rho}\hat{\sigma}_{gi}
-\hat{\sigma}_{gj}\hat{\sigma}_{gi}^{\dag}\hat{\rho}-\hat{\rho}\hat{\sigma}_{gj}\hat{\sigma}_{gi}^{\dag})], \label{eq6}\\
&\mathscr{L}_{ijc}\hat{\rho}=&\sum_{i,j=1,2}\frac{\gamma_{ijc}}{2}[(n_{ic}+1)(\hat{\sigma}_{\alpha i}\hat{\rho}\hat{\sigma}_{\alpha j}^{\dag}
+\hat{\sigma}_{\alpha j}\hat{\rho}\hat{\sigma}_{\alpha i}^{\dag}-\hat{\sigma}_{\alpha j}^{\dag}\hat{\sigma}_{\alpha i}\hat{\rho}-\hat{\rho}\hat{\sigma}_{\alpha j}^{\dag}\hat{\sigma}_{\alpha i})\nonumber\\
&&+n_{ic}(\hat{\sigma}_{\alpha i}^{\dag}\hat{\rho}\hat{\sigma}_{\alpha j}+\hat{\sigma}_{\alpha j}^{\dag}\hat{\rho}\hat{\sigma}_{\alpha i}
-\hat{\sigma}_{\alpha j}\hat{\sigma}_{\alpha i}^{\dag}\hat{\rho}-\hat{\rho}\hat{\sigma}_{\alpha j}\hat{\sigma}_{\alpha i}^{\dag})], \label{eq7}\\
&\mathscr{L}_{\Gamma_{C}}\hat{\rho}=&\frac{\Gamma_{C}}{2}[(N_{C}+1)(2\hat{\sigma}_{g\beta}\hat{\rho}\hat{\sigma}_{g\beta}^{\dag}-\hat{\sigma}_{g\beta}^{\dag}\hat{\sigma}_{g\beta}
\hat{\rho}-\hat{\rho}\hat{\sigma}_{g\beta}^{\dag}\hat{\sigma}_{g\beta})\nonumber\\
&&+N_{C}(2\hat{\sigma}_{\beta g}^{\dag}\hat{\rho}\hat{\sigma}_{\beta g}-\hat{\sigma}_{\beta g}\hat{\sigma}_{\beta g}^{\dag}\hat{\rho}-\hat{\rho}\hat{\sigma}_{\beta g}\hat{\sigma}_{\beta g}^{\dag})], \label{eq8}\\
&\mathscr{L}_{\Gamma}\hat{\rho}=&\frac{\Gamma}{2}(2\hat{\sigma}_{\beta\alpha}\hat{\rho}\hat{\sigma}_{\beta\alpha}^{\dag}-\hat{\sigma}_{\beta\alpha}^{\dag}\hat{\sigma}_{\beta\alpha}\hat{\rho}
-\hat{\rho}\hat{\sigma}_{\beta\alpha}^{\dag}\hat{\sigma}_{\beta\alpha}) \label{eq9}
\end{eqnarray}
\end{widetext}

\noindent  $\mathscr{L}_{ijh}\hat{\rho}$ in the expression (\ref{eq6}) denotes the dissipative effect of the ambient environment photon reservoirs.  And $\gamma_{iih}$=$\gamma_{ih}$, $\gamma_{jjh}$=$\gamma_{jh}$ are the spontaneous decay rates from the state \(|e_{i}\rangle(i$=$1,2)\) to the neutral ground state \(|g\rangle\) respectively. $\gamma_{ijh}$=$\gamma_{jih}$, the cross-coupling describes the effect of Fano interference. It is assumed that \(\gamma_{ijh}$=$\sqrt{\gamma_{ih}\gamma_{jh}}(i,j$=$1,2)\)  represents the maximal interference and \(\gamma_{ijh}$=$0\) for the minimal interference. And $n_{ih}(i$=$1,2)$ denotes the average electron occupations on the corresponding charge separation state.  The expression $\mathscr{L}_{ijc}\hat{\rho}$  describes the effect of the low temperature reservoirs with the average phonon numbers \(n_{ic}$=$[exp(\frac{(E_{e_{i}}-E_{\alpha})}{k_{B}T_{a}})-1]^{-1}(i$=$1,2)\) and \(T_{a}\) being the ambient temperature. $\gamma_{iic}$=$\gamma_{ic}$, $\gamma_{jjc}$=$\gamma_{jc}$ are the corresponding spontaneous decay rates from the state \(|e_{i}\rangle(i$=$1,2)\) to state \(|\alpha\rangle\). $\gamma_{ijc}$ is the cross-coupling that represents the effect of Fano interference, which is defined by $\gamma_{ijc}$=$\gamma_{jic}$$=$\(\gamma_{ijc}$=$\sqrt{\gamma_{ic}\gamma_{jc}}(i,j$=$1,2)\) with fully interference while \(\gamma_{ijc}=0\) for no interference. Similarly, $\mathscr{L}_{\Gamma_{C}}\hat{\rho}$ in expression (\ref{eq8}) describes another interaction between the PSII RC and ambient environment with $\hat{\sigma}_{g\beta}=|\beta\rangle\langle g|$, where \(N_{C}$=$[exp(\frac{(E_{\beta}-E_{g})}{k_{B}T_{a}})-1]^{-1}\) denotes the corresponding average phonon occupation numbers. The last term $\mathscr{L}_{\Gamma}\hat{\rho}$ describes a process that the PSII RC in state \(|\alpha\rangle\) decays to state $|\beta\rangle$. It leads to the photosynthetic power current proportional to the relaxation rate $\Gamma$ as defined before, and the operator $\hat{\sigma}_{\beta\alpha}$ is defined as $\hat{\sigma}_{\beta\alpha}$=$|\beta\rangle\langle\alpha|$. Thereupon, the elements of the density matrix $\rho$ can be written as follows,

\begin{widetext}
 \begin{align}
\dot{\rho}_{e_{1}e_{1}}={}&-\gamma_{1h}[(n_{1h}+1)\rho_{e_{1}e_{1}}-n_{1h}\rho_{gg}]-\gamma_{1c}[(n_{c}+1)\rho_{e_{1}e_{1}}-n_{c}\rho_{\alpha\alpha}]
                        \nonumber\\{}&-\gamma_{12h}[(n_{1h}+1) Re[\rho_{e_{1}e_{2}}]-\gamma_{12c}[(n_{c}+1) Re[\rho_{e_{1}e_{2}}],\nonumber\\
\dot{\rho}_{e_{2}e_{2}}={}&-\gamma_{2h}[(n_{2h}+1)\rho_{e_{2}e_{2}}-n_{1h}\rho_{gg}]-\gamma_{2c}[(n_{c}+1)\rho_{e_{2}e_{2}}-n_{c}\rho_{\alpha\alpha}]
                        \nonumber\\{}&-\gamma_{12h}[(n_{2h}+1)Re[\rho_{e_{1}e_{2}}]-\gamma_{12c}[(n_{c}+1) Re[\rho_{e_{1}e_{2}}],\nonumber\\
\dot{\rho}_{\alpha\alpha}={}&-\gamma_{1c}[(n_{c}+1)\rho_{e_{1}e_{1}}-n_{c}\rho_{\alpha\alpha}]+\gamma_{2c}[(n_{c}+1)\rho_{e_{2}e_{2}}-n_{c}\rho_{\alpha\alpha}]
                        \nonumber\\{}&+2\gamma_{12c}[(n_{c}+1)Re[\rho_{e_{1}e_{2}}]-\Gamma\rho_{\alpha\alpha},                               \label{eq10}\\
\dot{\rho}_{\beta\beta}={}&\Gamma\rho_{\alpha\alpha}-\Gamma_{C}[(\emph{N}_{C}+1)\rho_{\beta\beta}+\emph{N}_{C}\rho_{gg}],\nonumber\\
\dot{\rho}_{e_{1}e_{2}}={}&-\frac{\gamma_{1h}}{2}(n_{1h}+1)\rho_{e_{1}e_{2}}-\frac{\gamma_{2h}}{2}(n_{2h}+1)\rho_{e_{1}e_{2}}-\frac{\gamma_{12h}}{2}[(n_{1h}+1)\rho_{e_{1}e_{1}}+(n_{2h}+1)\rho_{e_{2}e_{2}}
              \nonumber\\{}&-n_{1h}\rho_{gg}-n_{2h}\rho_{gg}]-\frac{\gamma_{1c}+\gamma_{2c}}{2}(n_{c}+1)\rho_{e_{1}e_{2}}
              \nonumber\\{}&-\frac{\gamma_{12c}}{2}[(n_{c}+1)\rho_{e_{1}e_{1}}+(n_{c}+1)\rho_{e_{2}e_{2}}-2n_{c}\rho_{\alpha\alpha}].\nonumber
\end{align}
\end{widetext}

\noindent  where $\rho_{ii}$ describe the diagonal element and $\rho_{ij}$ is the non-diagonal element of the corresponding states. These equations provide a quantitative description of population dynamics in the PSII-RC, capturing the interplay between photon absorption, charge separation, and relaxation processes, forming the basis for subsequent LSTM modeling.

\subsection{Structure for LSTM }

\begin{table}
\begin{center}
\caption{Model parameters used in the numerical calculations.}
\label{Table1}
\vskip 0.2cm\setlength{\tabcolsep}{0.5cm}
\begin{tabular}{ccc}
\hline
                                                                        & Values                 & Units  \\
\hline
\(E_{1}-E_{\alpha}\)                                                    & 0.2                    & eV  \\
\(E_{\beta}-E_{g}\)                                                     & 0.2                    & eV  \\
\(\gamma_{1h}\)                                                         & 0.62                   & eV  \\
\(\gamma_{2h}\)                                                         & 0.56                   & eV  \\
\(\gamma_{1c}\)                                                         & 0.22                   & eV  \\
\(\gamma_{2c}\)                                                         & 0.82                   & eV  \\
\(\Gamma\)                                                              & 0.56                   & eV  \\
\(\Gamma_{C}\)                                                          & 0.68                   & eV  \\
\(n_{1h}\)                                                              & 60                     &     \\
\(n_{2h}\)                                                              & 58                     &     \\
\(T_{a}\)                                                               & 300                    & K   \\
\hline
\end{tabular}
\end{center}
\end{table}

Compared to traditional recurrent neural networks (RNNs)~\cite{shi2020analysis}, which suffer from gradient vanishing or explosion during backpropagation, long short-term memory (LSTM) networks\cite{halpern2020training} effectively address these issues by incorporating gated mechanisms that enhance the ability to capture long-term dependencies in sequential data.

The core architecture of LSTM \cite{wang2021software,smagulova2018memristor} consists of interconnected recurrent units, each containing specialized gate control units and a continuous memory cell. Each LSTM unit has three main gates: forget, input, and output gates. The forget gate determines the information to discard, the input gate updates the cell state, and the output gate controls the information passed to the next hidden state. The gate operations are expressed as:

\vskip -6pt\begin{align}
&f_{t} = \sigma(W_f \cdot [h_{t-1}, x_t] + b_f)\label{eq.11},\\
&i_{t} = \sigma(W_i \cdot [h_{t-1}, x_t] + b_i),\label{eq.12}\\
&\tilde{C}_t = \tanh(W_C \cdot [h_{t-1}, x_t] + b_C)\label{eq.13},\\
&o_t = \sigma(W_o \cdot [h_{t-1}, x_t] + b_o)\label{eq.14},
\end{align}
\begin{align}
&C_t = f_t \circ C_{t-1} + i_t \circ \tilde{C_t},\label{eq.15}\\
&h_t = o_t \circ \tanh(C_t)\label{eq.16}.
\end{align}
\begin{figure}
\centering
\subfigure[]{\includegraphics[width=0.48\columnwidth]{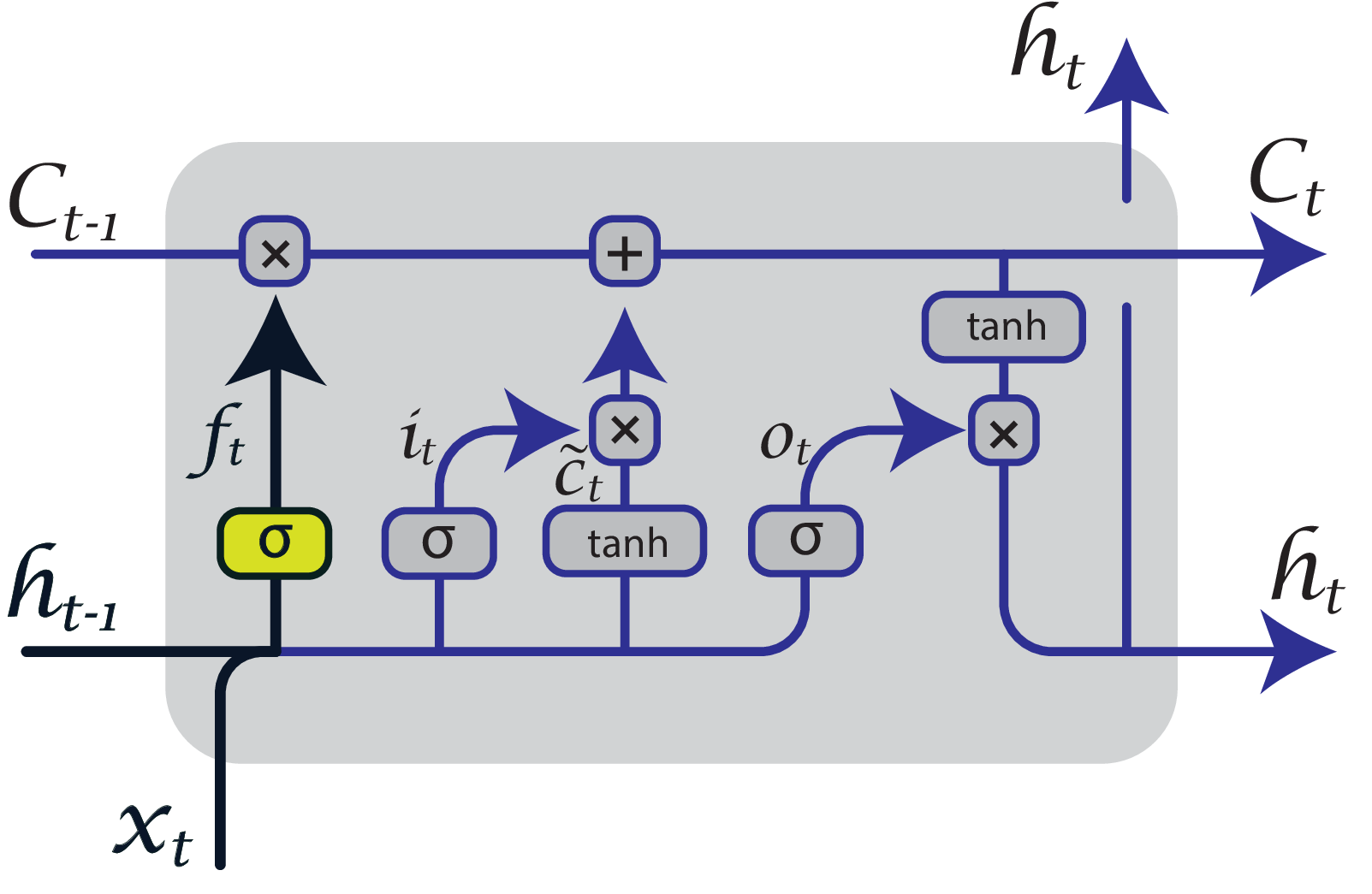}}
\vspace{-5pt}
\subfigure[]{\includegraphics[width=0.48\columnwidth]{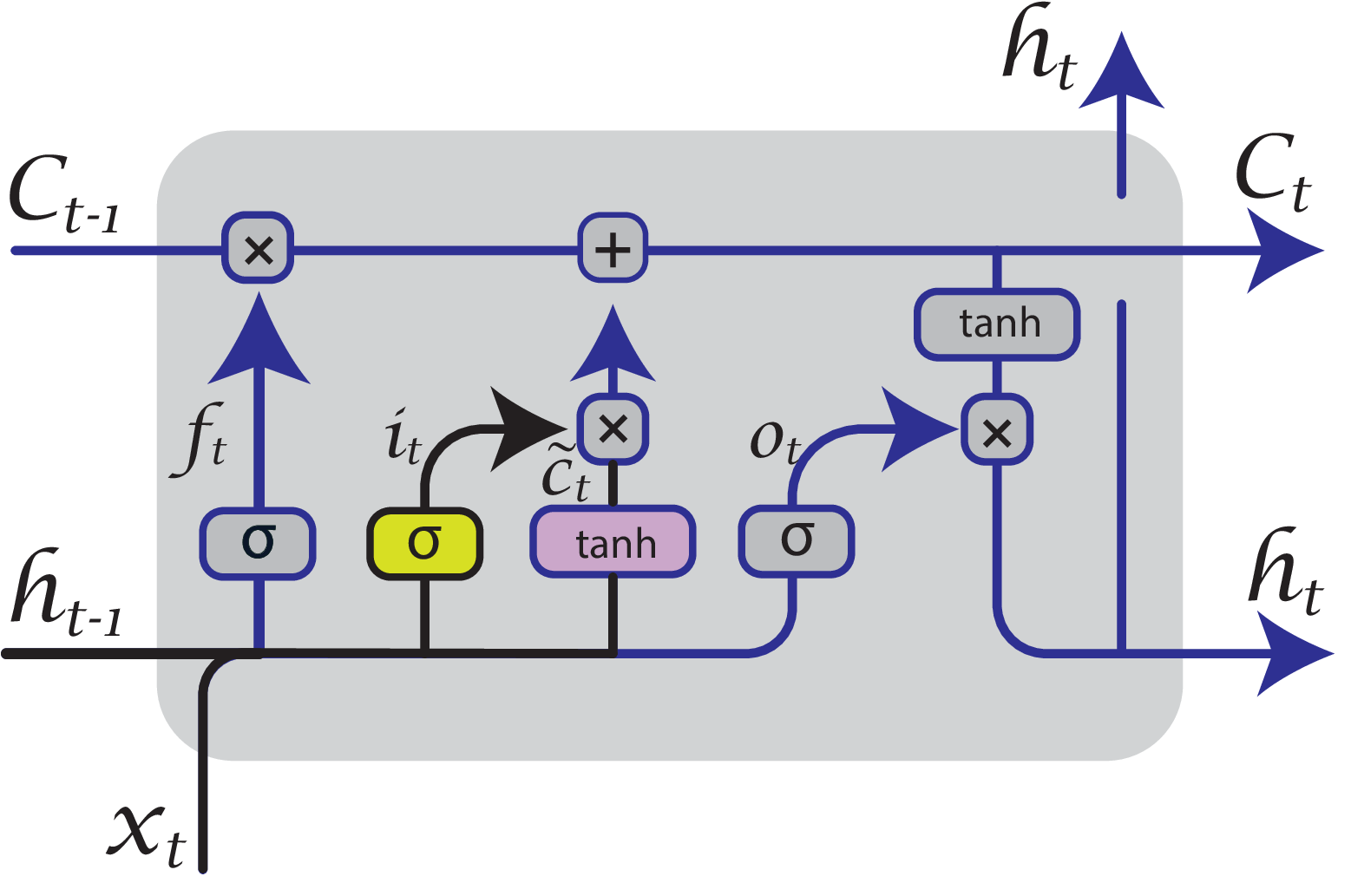}}
\vspace{-5pt}
\subfigure[]{\includegraphics[width=0.48\columnwidth]{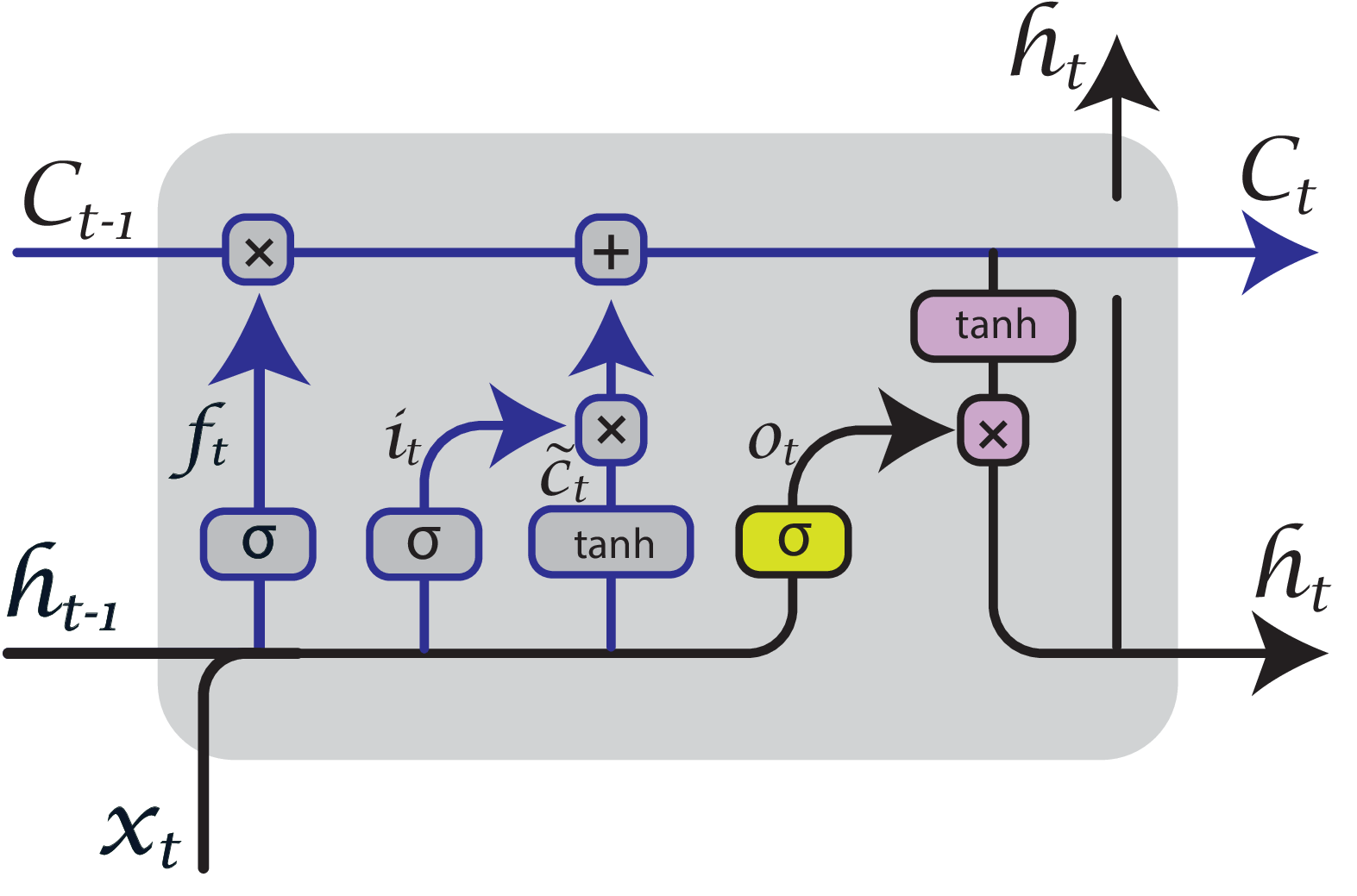}}
\vspace{-5pt}
\subfigure[]{\includegraphics[width=0.48\columnwidth]{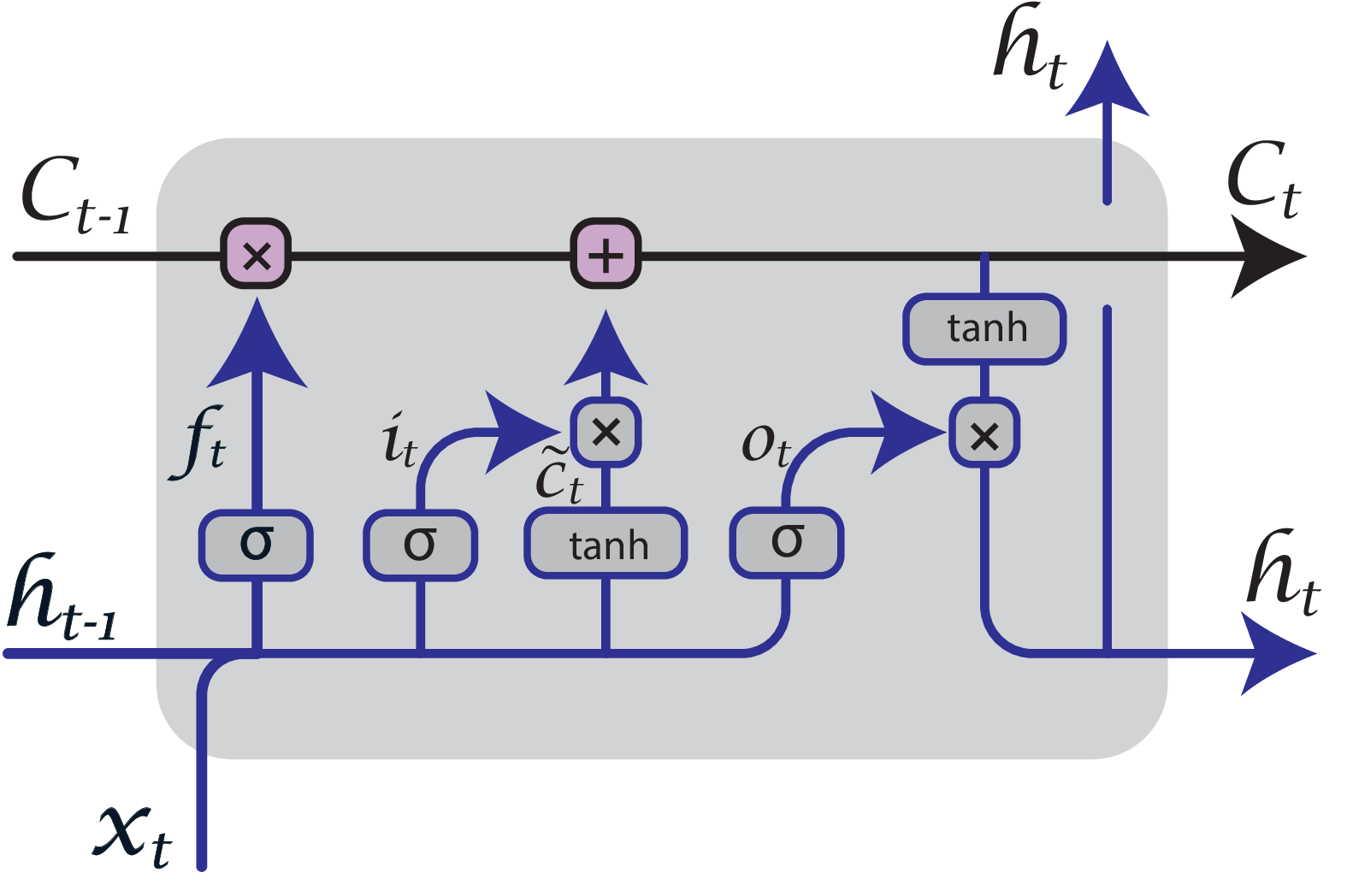}}
\vspace{-5pt}
\caption{The internal structures and operating mechanisms of (a) forget gate, (b) input gate, (c) output gate, (d) memory cell in LSTM networks.}\label{Fig.2}
\end{figure}
\vskip -6pt\noindent In LSTMs, the sigmoid function \(\sigma\) regulates the flow of information through each gate. The product of the weight matrix \([W_h, W_x]\) and the input vector\([h_{t-1}, x_t]\) , combined with a bias term \( b \) , determines the gate activations. During training, the network predicts through forward propagation, evaluates performance by comparing predictions with actual values \(\tilde{C_t}\), and updates weights using backpropagation \cite{kala2022forecasting}.

As shown in Fig. \ref{Fig.2}(a) and Eq. \ref{eq.11}, the forget gate takes the previous hidden state \( h_{t-1} \)  and current input \( x_t \) , passing them through a sigmoid function to decide how much information from the cell state \( C_{t-1} \) is retained or discarded. A value of 1 retains all information, while 0 discards it entirely. The input gate, shown in Fig. \ref{Fig.2}(b) and Eq. \ref{eq.13}, updates the cell state by selecting relevant values through a sigmoid function and generating candidate values \(C_t\) with a \(tanh\) function, combining them to update \(C_t\).

The output gate, illustrated in Fig. \ref{Fig.2}(c) and Eq. \ref{eq.14}, determines the next hidden state  \( h_t \)  by filtering the cell state through a combination of sigmoid and \(tanh\) activations. The memory unit $ C_t $ maintains or forgets information across multiple processing steps, regulated by these gating structures \cite{arunkumar2021forecasting}. This structure enables LSTMs to handle long-term dependencies, making them suitable for modeling population evolution in PSII-RC.

In this work, the Adam optimizer \cite{chang2019electricity,huang2023ultra} was employed to enhance training efficiency due to its adaptive learning rate and moment estimation features. Additionally, early stopping \cite{zhou2022carbon} and L2 regularization \cite{wang2022front,gao2023l2} with a coefficient of 0.01 were applied to prevent overfitting and reduce model complexity, ensuring stable long-term predictions.

\section{RESULTS AND DISCUSSIONS}
\subsection{Training the LSTM with error threshold training method}

In this work, a verifiable dataset was crucial to ensure the accuracy, reliability, and stability of the deep learning model. A dataset with ten thousand data points within 10 fs was generated using the Lindblad master equation (\ref{eq5}) and Eqs.~(\ref{eq10}), describing CT behavior via population dynamics in PSII-RC with model parameters listed in Table (\ref{Table1}). The data were split into training and test sets at an 80:20 ratio, with the first 8000 points used for training and the remaining 2000 for testing.

To assess the model, an innovative training approach was adopted. The trained model predicted 10 points, which were compared to actual values from the test set. A 0.0005 error threshold was applied: if all 10 predictions fell below this threshold, the model was saved; otherwise, retraining occurred. This threshold-based early stopping technique reduced training time and computational costs while ensuring stability and preventing overfitting. Unlike traditional methods optimizing overall loss, this approach prioritized high-precision predictions on specific test points, enhancing real-world performance.

\begin{figure}
\centering
\subfigure[]{\includegraphics[width=0.46\columnwidth]{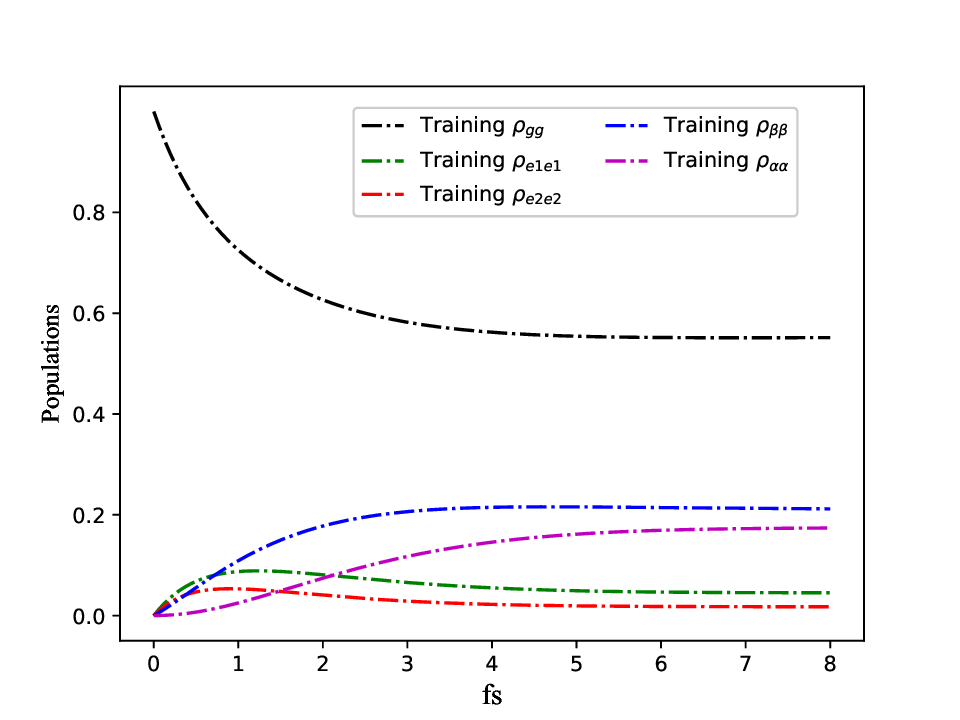}}\hspace{-12pt}
\subfigure[]{\includegraphics[width=0.50\columnwidth]{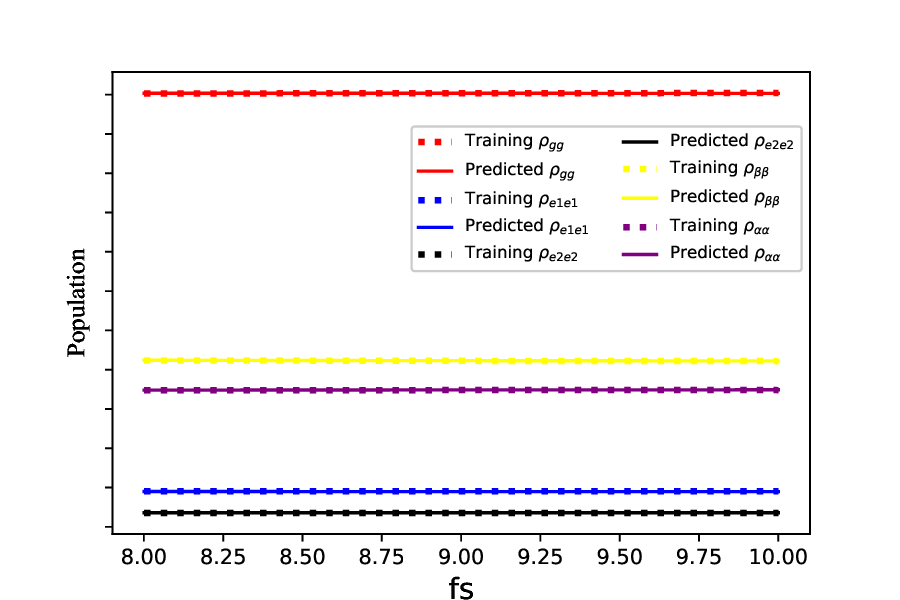}}
\vspace{-10pt}
\caption{(a) Training models fed by the actual data within [0, 8]fs for populations evolutions in the PSII-RC. (b) Predictive populations compared with the test data set within [8, 10]fs.}
\label{Fig.3}
\end{figure}

The LSTM was trained on data from [0, 8] fs using this method, and population evolutions in PSII-RC are shown in Fig. \ref{Fig.3}(a). To validate the model, populations were predicted within [8, 10] fs and compared with test data, as shown in Fig. \ref{Fig.3}(b). The high agreement between predicted and actual data confirms the model's long-term accuracy and stability, achieved through the optimization techniques described above.

\subsection{CT predicted by the LSTM}

\begin{figure}
\begin{flushleft}
\includegraphics[width=0.48\columnwidth]{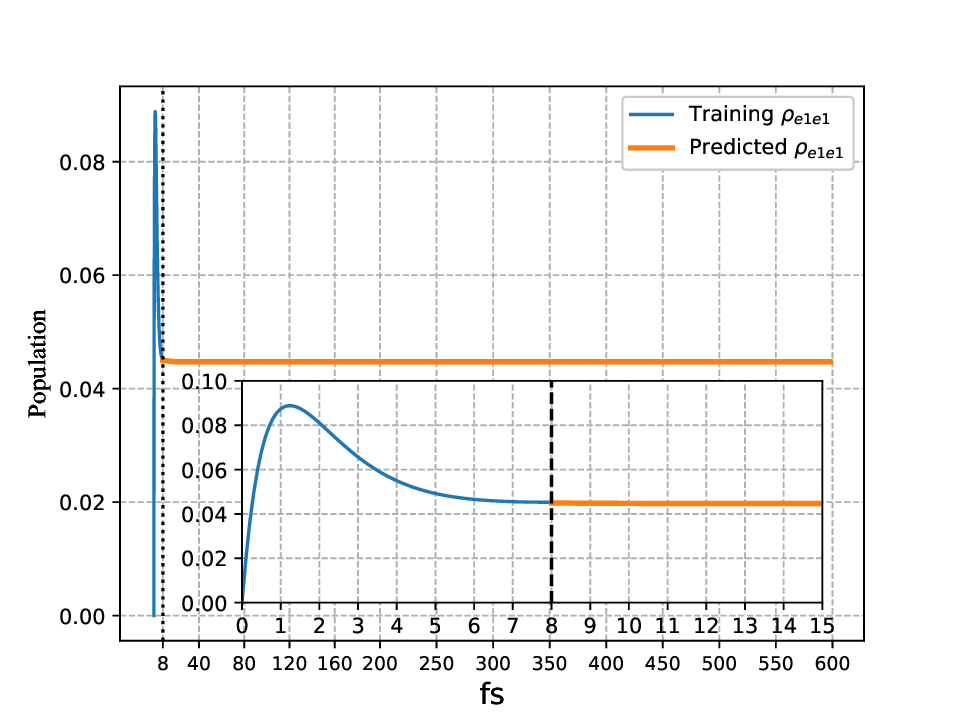}\vspace{-1pt}
\includegraphics[width=0.48\columnwidth]{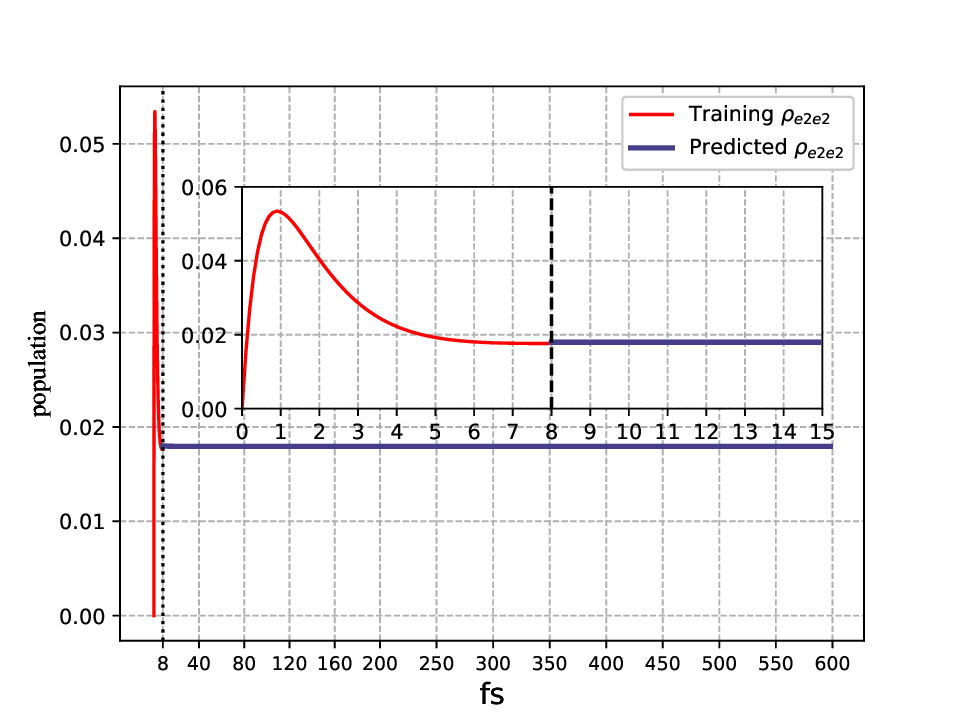}\vspace{-1pt}
\includegraphics[width=0.48\columnwidth]{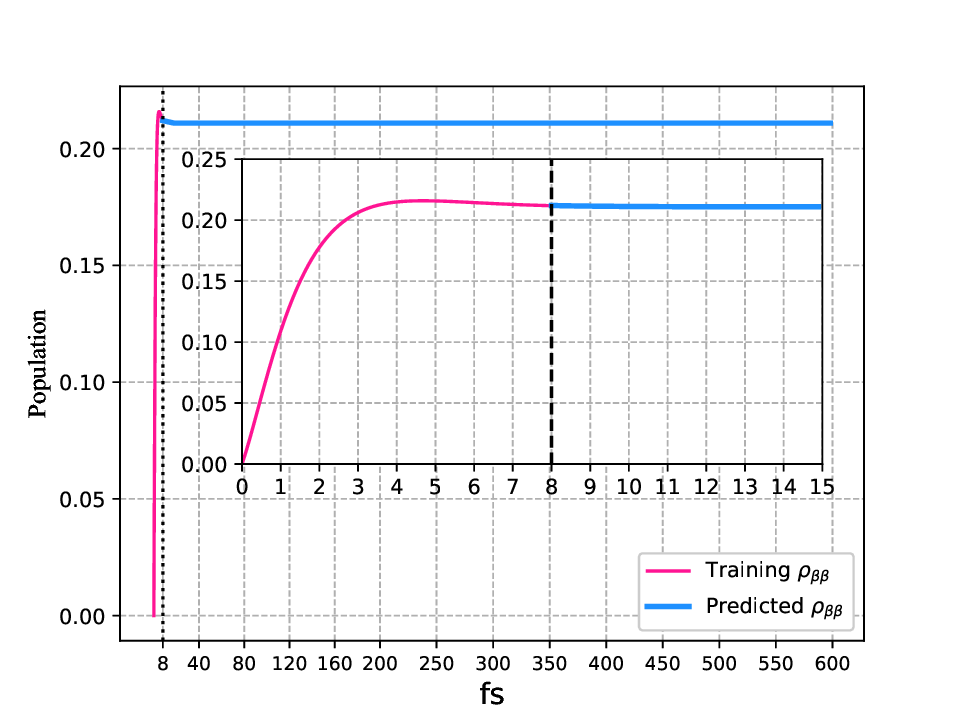}\vspace{-1pt}
\includegraphics[width=0.48\columnwidth]{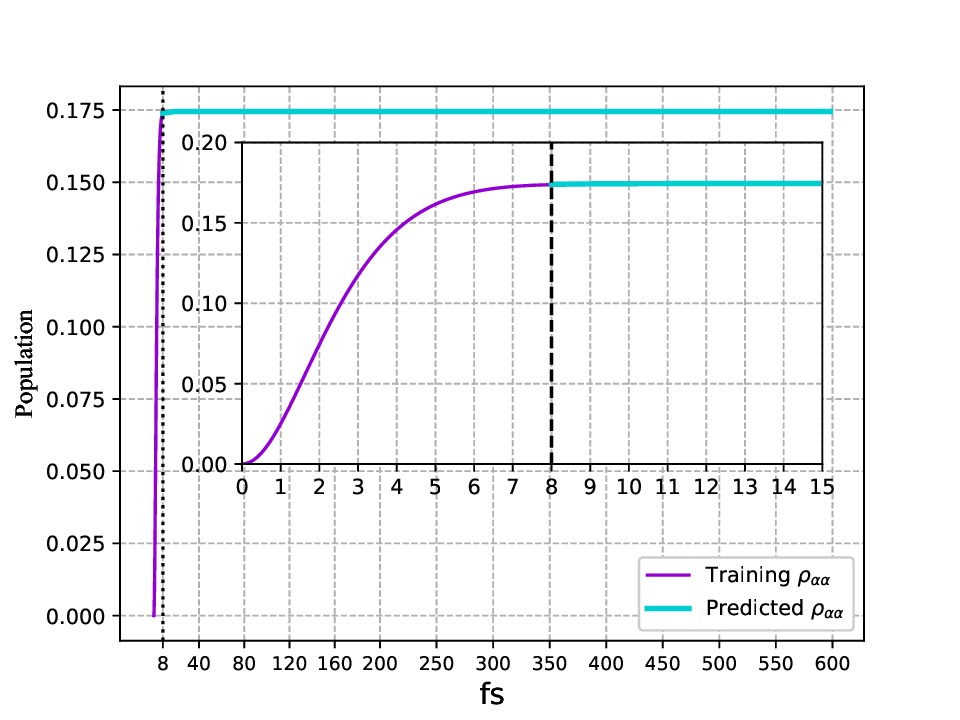}\vspace{-1pt}
\includegraphics[width=0.48\columnwidth]{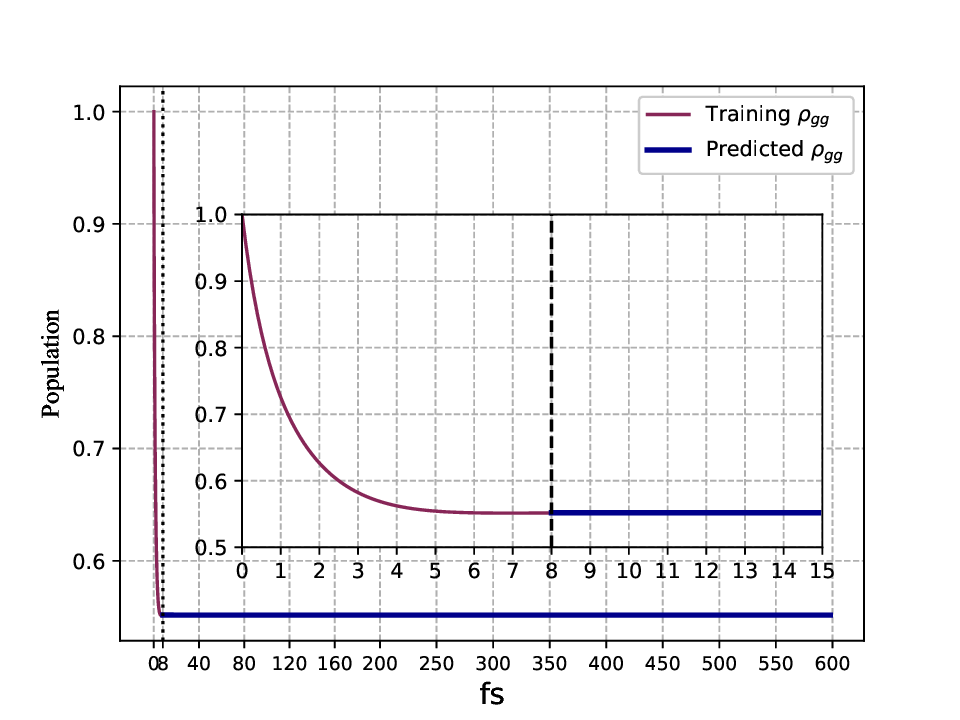}\vspace{-1pt}
\end{flushleft}\vspace{-10pt}
\caption{Population evolutions forecasted in [8, 600] fs by the proposed LSTM and training approach.}\label{Fig.4}
\end{figure}
In this section, we optimized the predictive performance by fine-tuning key hyperparameters, including optimizers, regularization, batch size, learning rate, dropout ratio, and early stopping strategies. Unlike previous works, we adopted the Adam optimizer with an initially low learning rate (0.00001) to ensure stable convergence, alongside L2 regularization (coefficient 0.01) and an extremely low dropout ratio (0.0001) to prevent overfitting while preserving learning capacity. Dynamic learning rate adjustments further enhanced performance at different training stages, improving CT process predictions.

A notable innovation was expanding the short temporal sequence of the training set by a factor of 10 to capture training patterns more effectively, then scaling back by the same factor during one-shot predictions, ensuring accuracy. A single-layer LSTM with 5-8 neurons, implemented in TensorFlow, was employed to model the CT process \cite{del2021learning}. After 100 iterations, the model automatically saved its state once stability within a 0.0005 prediction error threshold was achieved, significantly enhancing training efficiency and reducing computational costs(\href{https://pan.baidu.com/s/1nbQsKLFoplun5AaTJOON1A?pwd=zzrz}{Original codes in ``Model training and prediction"}).

The results confirmed the effectiveness of these strategies. As training progressed, the model demonstrated remarkable generalization for CT predictions beyond the training set. In Fig. \ref{Fig.4}, the vertical black line at 8 fs separates training data from predictions, with curves on the right side representing outcomes from progressively tuned hyperparameters. The inset highlights near-perfect coincidence at 8 fs, validating prediction accuracy(\href{https://pan.baidu.com/s/1nbQsKLFoplun5AaTJOON1A?pwd=zzrz}{Original codes in ``Plot figures"}).

Interestingly, population evolution over time displayed horizontal line distributions (Fig. \ref{Fig.4}), implying consistent predictive performance across extended periods. Comparing the predicted and actual values at 600 fs (Table \ref{Table2}) revealed differences on the order of $10^{-4}$, underscoring the model's high precision over long timescales. These results highlight the model¡¯s robustness and its potential for accurately predicting long-term quantum dynamics in complex systems.

\begin{table}
\begin{center}
\caption{{\scriptsize Comparison of actual and predicted values at 600 fs.}}
\vskip 0.1cm\setlength{\tabcolsep}{0.1cm}
\begin{tabular}{|c|c|c|c|}
\hline
 Populations & Actual value & Predicted value   & Difference \\
\hline
\(\rho_{e1e1}\)           &0.04543156    & 0.04473562    &  0.00069594 \\
\(\rho_{e2e2}\)           &0.0177977     & 0.017938      & -0.00014031 \\
\(\rho_{\alpha\alpha}\)   &0.17367974    & 0.17446348    & -0.00078374 \\
\(\rho_{\beta\beta}\)     &0.21068688    & 0.21092759    & -0.00024071 \\
\(\rho_{gg}\)             &0.55240412    & 0.5517195     &  0.00068464 \\
\hline
\end{tabular}\label{Table2}
\end{center}
\end{table}

\section{CONCLUSION AND OUTLOOK}

In conclusion, this work employed an LSTM network with an error-threshold training method to predict charge transport (CT) in Photosystem II (PSII). Using 8000 data points generated by the Lindblad master equation, the model was trained with a 0.0005 error threshold, ensuring high prediction accuracy. The results demonstrate that the LSTM achieved long-term predictions with a precision of $10^{-4}$ magnitude from 8 fs to 600 fs, effectively enhancing the accuracy of long-term series predictions. Proper hyperparameter tuning, including a dynamically adjusted learning rate and regularization strategies, significantly improved the model¡¯s stability and reliability. This study not only advances CT prediction in PSII but also provides robust methodological support for simulating complex physical processes, paving the way for further applications in quantum dynamics.

\section*{Author contributions}

S. C. Zhao conceived the idea. Z. R. Zhao performed the numerical computations and wrote the draft, and S. C. Zhao did the analysis and revised the paper. Y. M. Huang participated in part of the discussion.

\section{Acknowledgment}

This work is supported by the National Natural Science Foundation of China ( Grant Nos. 62065009 and 61565008 ), General Program of Yunnan Applied Basic Research Project, China ( Grant No. 2016FB009 ).
\section*{Data Availability Statement}

This manuscript has associated data in a data repository.[Authors' comment: All data included in this manuscript are available upon resonable request by contacting with the corresponding author.] The Supporting Information is available free of charge at: \href{https://pan.baidu.com/s/1nbQsKLFoplun5AaTJOON1A?pwd=zzrz} {Original codes in Supplement information}.

\section*{Conflict of Interest}

The authors declare that they have no conflict of interest. This article does not contain any studies with human participants or animals performed by any of the authors. Informed consent was obtained from all individual participants included in the study.
\bibliography{reference}
\bibliographystyle{apsrev4-1}
\end{document}